\documentclass{elsart}

\usepackage{cite}
\usepackage{dcolumn}
\usepackage{graphicx}

\begin{document}

\begin{frontmatter}

\title{Local approximation to the critical parameters of quantum wells}

\author{Francisco M Fern\'andez\thanksref{FMF}} and
\author{Javier Garcia}

\address{INIFTA (UNLP, CCT La Plata-CONICET), Divisi\'on Qu\'imica Te\'orica,
Blvd. 113 S/N,  Sucursal 4, Casilla de Correo 16, 1900 La Plata,
Argentina}

\thanks[FMF]{e--mail: fernande@quimica.unlp.edu.ar}

\begin{abstract}
We calculate the critical parameters for some simple quantum wells
by means of the Riccati-Pad\'{e} method. The original approach
converges reasonably well for nonzero angular-momentum quantum
number $l$ but rather too slowly for the s states. We therefore
propose a simple modification that yields remarkably accurate
results for the latter case. The rate of convergence of both
methods increases with $l$ and decreases with the radial quantum
number $n$. We compare RPM results with WKB ones for sufficiently
large values of $l$. As illustrative examples we choose the
one-dimensional and central-field Gaussian wells as well as the
Yukawa potential. The application of perturbation theory by means
of the RPM to a class of rational potentials yields interesting
and baffling unphysical results.
\end{abstract}

\end{frontmatter}

\section{Introduction}

\label{sec:intro}

The accurate calculation of the number of bound states supported by a finite
quantum-mechanical potential well is of great physical and mathematical
importance and for this reason there has been considerable interest in the
derivation of upper and lower bounds\cite
{B52,S61,GR65,C65a,C65b,N83,CMS95,CKMS96,CK97,BC03a,BC03b,BC04,M06}. Most of
those bounds are given in terms of the potential-energy function. In a
recent paper Liverts and Barnea\cite{LB11} proceeded in a different way and
proposed the calculation of the critical parameters for negative
central-field quantum wells. To this end they applied two exact methods and
the WKB approach, the latter for the estimation of the large-quantum number
behaviour of the critical parameters. In this context a critical parameter
is the value of a potential parameter for which an energy eigenvalue is
exactly zero (what the authors call a transition state). As they pointed
out, one can obtain the exact number of bound states from the tables of
critical parameters, as well as other relevant information about the
eigenvalue equation\cite{LB11}.

There are local and global methods for the calculation of eigenvalues and
eigenfunctions. The former are based on the behaviour of the solution at a
properly chosen coordinate point; for example, a power-series expansion. On
the other hand, global approaches like the variational method take into
account the whole coordinate interval (through expectation values of the
associated linear operators, etc.). In principle, local methods are expected
to be unsuitable for the calculation of critical parameters. Even the
Riccati-Pad\'{e} method (RPM)\cite{FMT89a,FMT89b}, based on Pad\'{e}
approximants, was shown to be impractical for the calculation of the
eigenvalues of the Yukawa potential close to the zero-energy threshold
(transition state)\cite{FMT89a}. The purpose of this paper is to investigate
in more detail whether those earlier results already prove that the RPM is
actually useless for the calculation of critical parameters.

In section \ref{sec:RPM} we outline the main ideas of the RPM. In section
\ref{sec:wells} we briefly discuss the solutions of the Schr\"{o}dinger
equation with even-parity potential wells. In section \ref{sec:examples} we
apply the approach to some simple one-dimensional models: the
P\"oschl-Teller potential, the Gaussian well and a rational potential. We
calculate some critical parameters and the corresponding eigenfunctions for
the first and third cases. In section \ref{sec:MRPM} we propose a modified
version of the RPM that is more suitable for the calculation of critical
parameters and apply it to the Gaussian well. Some of those results also
apply to s-states of the analogous central-field model. In section \ref
{sec:central-field} we apply the modified RPM to the s-states of
central-field models and choose the Yukawa potential as a suitable
illustrative example. We also show that the original RPM is suitable for the
calculation of critical parameters of states with $l>0$. In section \ref
{sec:PT} we discuss the application of perturbation theory to a model with a
rational potential that is exactly solvable at threshold. In this case we
discuss the appearance of spurious RPM eigenvalues. Finally, in section \ref
{sec:conclusions} we summarize the main results and draw conclusions.

\section{The Riccati-Pad\'{e} method}

\label{sec:RPM}

We consider the eigenvalue equation
\begin{equation}
\psi ^{\prime \prime }(x)+Q(E,x)\psi (x)=0,\,-\infty <x<\infty ,
\label{eq:eigen_eq}
\end{equation}
were $E$ is the eigenvalue. We assume that $\psi (x)$ can be expanded about
the origin as
\begin{equation}
\psi (x)=x^{s}\sum_{j=0}^{\infty }c_{j}x^{\beta j},\,\alpha ,\beta >0.
\end{equation}
It is clear that
\begin{equation}
f(x)=\frac{s}{x}-\frac{\psi ^{\prime }(x)}{\psi (x)},
\end{equation}
can be expanded about the origin as
\begin{equation}
f(x)=x^{\beta -1}\sum_{j=0}^{\infty }f_{j}z^{j},\,z=x^{\beta }.
\end{equation}

We approximate $f(x)$ by means of a rational function of the form $x^{\beta
-1}[M/N](z)$ where
\begin{equation}
\lbrack M/N](z)=\frac{\sum_{j=0}^{M}a_{j}z^{j}}{\sum_{j=0}^{N}b_{j}z^{j}}%
=T(M+N+1,z)+O(z^{M+N+2}) ,  \label{eq:[M/N](z)_cond}
\end{equation}
and
\begin{equation}
T(n,z)=\sum_{j=0}^{n}f_{j}z^{j}.
\end{equation}
We choose $M\geq N$ and define $d=M-N$. It is not possible to satisfy the
condition (\ref{eq:[M/N](z)_cond}) unless $H_{D}^{d}=\left|
f_{i+j+d-1}\right| _{i,j=1}^{D}=0$, $D=N+1$. The coefficients $f_{j}$, and
thereby the Hankel determinant $H_{D}^{d}$, depend on the eigenvalue $E$.
Some of the roots $E^{[D,d]}$ of $H_{D}^{d}(E)=0$ converge toward the
eigenvalues of Eq.~(\ref{eq:eigen_eq} ) as $D$ increases\cite{FMT89a,FMT89b}.

The ordinary Pad\'{e} approximation to $f(x)$ is
\begin{equation}
\lbrack M/N](z)=\frac{\sum_{j=0}^{M}a_{j}z^{j}}{\sum_{j=0}^{N}b_{j}z^{j}}%
=T(M+N,z)+O(z^{M+N+1}) .  \label{eq:[M/N](z)_Pade}
\end{equation}
If $z_{0}$ is a zero of the denominator then
\begin{equation}
z_{0}=-\frac{b_{N-1}}{b_{N}}-\frac{b_{N-2}}{b_{N}z_{0}}-\ldots -\frac{b_{0}}{%
b_{N}z_{0}^{N-1}} .  \label{eq:z0}
\end{equation}
Suppose that $E^{*}$ is a root of $b_{N}(E)=0$ and that $b_{N-1}(E)$ does
not vanish in the interval $(E^{*}-\epsilon ,E^{*}+\epsilon )$ for a
sufficiently small positive real number $\epsilon$. Therefore $%
|z_{0}|\rightarrow \infty $ as $E\rightarrow E^{*}$. The coefficient $b_{N}$
is proportional to the Hankel determinant $H_{N-1}^{d+1}(E)$ so that the
Hankel condition $H_{D}^{d}(E)=0$ is equivalent to moving a singularity of a
rational approximation towards infinity\cite{AB11}. It is also equivalent to
moving a zero of the approximate $\psi (x)$ towards infinity. Consequently,
it appears to be reasonable to assume that the Hankel condition is
equivalent to selecting bound states that vanish at infinity.

The strategy just outlined applies to other nonlinear equations and for this
reason Amore and Fern\'andez\cite{AF11} chose the more general name
Pad\'e-Hankel method which was later discussed by Abbasbandy and Bervillier%
\cite{AB11}. However, for historical reasons we prefer to keep the original
name RPM when the problem is a Riccati equation derived from the
Schr\"odinger one\cite{FMT89a,FMT89b}.

\section{Parity-invariant finite wells}

\label{sec:wells}

The Riccati-Pad\'{e} method is known to produce accurate eigenvalues for
infinite wells or sufficiently deep finite ones\cite{FMT89a,FMT89b}. The
purpose of this section is to investigate to which extent it is possible to
apply the RPM to shallow wells. To this end we consider the eigenvalue
equation (\ref{eq:eigen_eq}) with
\begin{equation}
Q(E,x)=2\left[ E-V(x)\right],
\end{equation}
where the potential-energy function $V(x)$ exhibits a minimum at $V(0)<0$
and $V(x\rightarrow \pm \infty )=0$. In order to simplify the discussion we
assume that $V(-x)=V(x)$ so that
\begin{equation}
Q(E,x)=\sum_{j=0}^{\infty }Q_{j}x^{2j}.
\end{equation}
The shape of a parity-invariant potential is commonly determined by a
smaller number of parameters. In addition to it, the results for the odd
states also apply to the solutions of the Schr\"{o}dinger equation with a
central-field potential having zero angular momentum quantum number $l$ (s
states).

The eigenfunctions of the Hamiltonian operator with a parity-invariant
potential are even or odd; therefore, $s=0$ for the former, $s=1$ for the
latter and $\beta =2$ in both cases. We thus have
\begin{equation}
f(x)=x\sum_{j=0}^{\infty }f_{j}x^{2j},
\end{equation}
where the first coefficients are

\begin{eqnarray}
f_{0} &=&\frac{Q_{0}}{1+2s} ,  \nonumber \\
f_{1} &=&\frac{Q_{0}^{2}}{(3+2s)(1+2s)^{2}}+\frac{Q_{1}}{3+2s} ,  \nonumber
\\
f_{2} &=&\frac{2Q_{0}^{3}}{(5+2s)(3+2s)(1+2s)^{3}}+\frac{2Q_{0}Q_{1}}{%
(5+2s)(1+2s)(3+2s)}+\frac{Q_{2}}{(5+2s)} .  \nonumber \\
&&
\end{eqnarray}
Besides, the function $f(x)$ is a solution to the Riccati equation
\begin{equation}
f^{\prime }(x)+\frac{2s}{x}f(x)-f(x)^{2}-Q(E,x)=0.  \label{eq:Riccati}
\end{equation}

For convenience we define $v_{0}=-V(0)>0$ and $v(x)=-V(x)/v_{0}$. For all
values of the well depth $v_{0}>0$ there is always a bound ground state with
energy $E_{0}$. The number of bound states with energies $E_{0}<E_{1}<\ldots
<E_{n}$ $<0$ depends on $v_{0}$. As $v_{0}$ decreases the highest
bound-state energy, say $E_{n}$, approaches the threshold $E=0$ from below
and we have a critical well parameter $v_{0,n}$ when $E_{n}=0$.
Consequently, there are $n+1$ bound states when $v_{0,n}<v_0<v_{0,n+1}$.

We assume that $V(x)$ vanishes faster than $x^{-2}$ as $|x|\rightarrow
\infty $; that is to say
\begin{equation}
\lim_{|x|\rightarrow \infty }x^{2}V(x)=0 .  \label{eq:V(x)_asympt}
\end{equation}
Therefore, an eigenfunction for the arbitrary energy $E<0$ behaves
asymptotically as
\begin{equation}
\psi (x)\sim A(E,v_{0})e^{-\alpha x}+B(E,v_{0})e^{\alpha x},\,|x|\rightarrow
\infty ,  \label{eq:psi_asympt_1}
\end{equation}
where $\alpha =\sqrt{-2E}$. The bound states are given by the condition $%
B(E_{j},v_{0})=0$ that leads to square-integrable eigenfunctions.

When $E=0$ the two asymptotic solutions in Eq.~(\ref{eq:psi_asympt_1}) are
linearly dependent. In this case the general solution to $\psi ^{\prime
\prime }(x)=0$ behaves as
\begin{equation}
\psi (x)\sim A(v_{0})+B(v_{0})x,\,E=0,\,|x|\rightarrow \infty .
\label{eq:psi_asympt_2}
\end{equation}
The solution at threshold is not square integrable but we can think of it as
the limit of a square integrable one $\lim_{E\rightarrow
0}A(E,v_{0})e^{-\alpha x}=A(v_{0})$ ($|x|\rightarrow \infty $). Therefore,
the critical parameters are roots of $B(v_{0,n})=0$ and the boundary
condition at threshold is
\begin{equation}
\lim_{|x|\rightarrow \infty }\psi ^{\prime }(x)=0,\,E=0 .  \label{eq:BC_E=0}
\end{equation}

\section{Examples}

\label{sec:examples}

In what follows we discuss some simple model potentials to illustrate the
application of the RPM.

\subsection{Modified P\"{o}schl-Teller potential}

\label{subsec:MPT}

As a first example we consider the modified P\"{o}schl-Teller potential
\begin{equation}
V(x)=-\frac{v_{0}}{\cosh ^{2}(x)},\,v_{0}>0 .  \label{eq:V_MPT}
\end{equation}
There are two reasons for this choice: first, we can solve the
Schr\"{o}dinger equation and obtain a simple expression for the eigenvalues:%
\cite{F99}
\begin{eqnarray}
E_{n} &=&-\frac{(\lambda -n-1)^{2}}{2},\,n=0,1,\ldots  \nonumber \\
\lambda &=&\frac{1+\sqrt{1+8v_{0}}}{2} .  \label{eq:En_MPT}
\end{eqnarray}
Second, the exact bound-state solutions are hypergeometric functions of $%
y=\cosh ^{2}x$ so that the RPM based on an $x$-power series can only yield
approximate results. Therefore, this model is a suitable benchmark for
testing the performance of the approach.

As discussed above the critical values of the potential parameter $v_{0}$
are determined by the condition $E_{n}(v_{0,n})=0$. It follows from equation
(\ref{eq:En_MPT}) that $v_{0,n}=n(n+1)/2$, $n=0,1,\ldots $. We first
investigate if there are converging sequences of roots $E^{[D,d]}$ of the
Hankel determinant $H_{D}^{d}$ as $v_{0}$ approaches $v_{0,0}=0$ and $%
v_{0,2}=3$. We calculated $E^{[D,0]}$, $D=2,3,\ldots $ for $%
v_{0}=v_{0,n}+10^{-k}$, for $k=1,2,\ldots $. The results show that there are
convergent sequences of roots for $D=D_{k},D_{k}+1,\ldots $ and that the
starting point of each sequence $D_{k}$ increases as $k$ increases. There
seems to be convergent sequences no matter how large the value of $k$. In
other words, the RPM appears to be successful no matter how close $v_{0}$ is
to the critical value $v_{0,n}$.

Since the roots of the Hankel determinants $H_{D}^{d}(E,v_{0})=0$ give rise
to sequences that clearly converge towards the eigenvalues $E_{n}(v_{0}) $
for $v_{0}$ quite close to $v_{0,n}$, then one would expect to find
sequences of roots of $H_{D}^{d}(E=0,v_{0})=0$ that converge towards the
critical parameters $v_{0,n}$. This is exactly the case for this model and
one obtains the critical parameters with any desired degree of accuracy with
Hankel determinants of relatively small dimension. There are, however, two
surprising facts. The first one is that the RPM yields all the critical
parameters $v_{0,n}$, $n=1,2,\ldots $ when choosing either the even ($s=0$)
or odd ($s=1$) functions. One would expect the even or odd values of $n$ to
appear separately with even or odd functions, respectively. The second
surprising fact is that the RPM with $s=0$ yields the critical parameters
with odd $n$ more accurately than those with even $n$. The opposite
situation takes place when choosing $s=1$.

We can understand the occurrence of twice as much critical parameters as
expected by obtaining the corresponding wave functions in the usual way\cite
{F99}. If $\psi _{n,s}(x)$ denotes the solution of parity $s$ for $%
v_{0}=v_{0,n}$ and $E=0$, then the first of them are given by
\begin{eqnarray}
\psi _{1,0}(x) &=&\frac{2x}{e^{2x}+1}-x+1 ,  \nonumber \\
\psi _{1,1}(x) &=&1-\frac{2}{e^{2x}+1} ,  \nonumber \\
\psi _{2,0}(x) &=&\frac{2\left( 4e^{2x}-e^{4x}-1\right) }{e^{4x}+2e^{2x}+1} ,
\nonumber \\
\psi _{2,1}(x) &=&-\frac{\left[ e^{4x}\left( 2x-3\right)
-8xe^{2x}+2x+3\right] }{4\left( e^{4x}+2e^{2x}+1\right) } .
\label{eq:psi_MPT}
\end{eqnarray}
We appreciate that $\psi _{1,1}(x)$ and $\psi _{2,0}(x)$ are convergent
while $\psi _{1,0}(x)$ and $\psi _{2,1}(x)$ are divergent. In general, $\psi
_{n,s}$ is convergent or divergent provided that $n+s$ is even or odd,
respectively:
\begin{eqnarray}
\lim_{|x|\rightarrow \infty }\psi _{n,s,}(x) &=&A,\,n+s=2k ,  \nonumber \\
\lim_{|x|\rightarrow \infty }x^{-1}\psi _{n,s}(x) &=&B,\,n+s=2k-1 ,
\nonumber \\
k &=&1,2,\ldots  \label{eq:lim_psi_ns}
\end{eqnarray}
We conclude that the RPM approaches both the convergent and divergent
solutions for this problem when $E=0$. This is the reason why the whole set
of critical parameters $v_{0,n}$ appears for both the even and odd
solutions: half of them are convergent and the other half divergent. It is
clear, as already argued above, that the RPM does not yield the exact result
because the exact $f(x)$ is not a rational function of $x$ for any of the
functions~(\ref{eq:psi_MPT}).

It is not clear to us why the RPM with $s=0$ ($s=1$) yields the critical
parameters with odd (even) $n$ more accurately. We will discuss this point
with somewhat more detail below by means of a solvable model with a rational
potential.

\subsection{Gaussian well}

\label{subsec:Gaussian}

The Gaussian well
\begin{equation}
V(x)=-v_{0}e^{-x^{2}} ,  \label{eq:V_Gauss}
\end{equation}
is another suitable choice because the potential is extremely simple but the
Schr\"{o}dinger equation is not exactly solvable. In this case the behaviour
of the sequences of roots of $H_{D}^{d}(E,v_{0})=0$ appears to be similar
except that the starting point $D_{k}$ of a given sequence increases more
pronouncedly as $v_{0}$ approaches $v_{0,n}$ and we could not find
converging sequences of roots of $H_{D}^{d}(E=0,v_{0})=0$. It is not clear
to us which is the feature of this well that makes such a difference. Since
the present form of the RPM appears to be unsuitable for obtaining the
critical parameters for this problem in section~\ref{sec:MRPM} we will
discuss an improved version of the approach.

\subsection{Rational potential}

\label{subsec:RatPot}

The third example in this section is the potential well
\begin{equation}
V(x)=-\frac{v_{0}}{\left( 1+x^{2}\right) ^{2}} ,  \label{eq:V_rational}
\end{equation}
that satisfies the condition (\ref{eq:V(x)_asympt}). Joseph\cite{J67b}
studied the family of central-field potentials $V(r)=-\lambda r^{\alpha
-2}(r_{0}^{2}+r^{2})^{-\alpha }$ in his discussion of local degeneracy.
Clearly, the potential (\ref{eq:V_rational}) is the one-dimensional version
of the case $\alpha =2$. Besides, present results for the odd states should
agree with those obtained by Joseph for $\alpha =2$ and $l=0$.

The roots of the Hankel determinants $H_{D}^{0}(E=0,v_{0})$ yield exact
critical parameters $v_{0,n}=n(n+2)/2$, $n=1,2,\ldots $. These results
correspond to exact rational solutions to the Riccati equation (\ref
{eq:Riccati}). In order to understand their meaning we construct the
corresponding wavefunctions as
\begin{equation}
\psi _{n,s}=x^{s}\exp \left[ -\int f(x)\,dx\right].
\end{equation}
The first even and odd ones are

\begin{eqnarray}
\psi _{1,0}(x) &=&\frac{1-x^{2}}{\sqrt{x^{2}+1}} ,  \nonumber \\
\psi _{2,0}(x) &=&\frac{1-3x^{2}}{x^{2}+1} ,  \nonumber \\
\psi _{3,0}(x) &=&\frac{\left( x^{2}+2x-1\right) \left( x^{2}-2x-1\right) }{%
\left( 1+x^{2}\right) ^{3/2}} ,  \nonumber \\
\psi _{4,0}(x) &=&\frac{5x^{4}-10x^{2}+1}{\left( 1+x^{2}\right) ^{2}} ,
\label{eq:psi_rat_even}
\end{eqnarray}
and

\begin{eqnarray}
\psi _{1,1}(x) &=&\frac{x}{\sqrt{1+x^{2}}} ,  \nonumber \\
\psi _{2,1}(x) &=&\frac{x\left( x^{2}-3\right) }{1+x^{2}} ,  \nonumber \\
\psi _{3,1}(x) &=&\frac{x\left( x^{2}-1\right) }{\left( 1+x^{2}\right) ^{3/2}%
} ,  \nonumber \\
\psi _{4,1}(x) &=&\frac{x\left( x^{4}-10x^{2}+5\right) }{\left(
1+x^{2}\right) ^{2}} ,  \label{eq:psi_rat_odd}
\end{eqnarray}
respectively. As in the case of the modified P\"{o}schl-Teller potential the
solutions $\psi _{n,s}$ are convergent or divergent provided that $n+s$ is
even or odd, respectively; more precisely, they satisfy equations~(\ref
{eq:lim_psi_ns}). According to the discussion of section~\ref{sec:wells} $%
\psi _{2k,0}(x)$ are the even solutions to the Schr\"{o}dinger equation for $%
E=0$ and $v_{0}=2k(k+1)$ (satisfy the condition $B(v_{0})=0$). On the other
hand, $\psi _{2k-1,1}(x)$ are the odd solutions for $E=0$ and $%
v_{0}=(4k^{2}-1)/2$ , $k=1,2,\ldots $. The latter agree with Joseph's ones
when $\lambda =2v_{0}$\cite{J67b}. The remaining solutions $\psi _{2k,1}(x)$
and $\psi _{2k-1,0}(x) $ are the unphysical divergent solutions to the
Schr\"{o}dinger equation. We see that the RPM yields the exact convergent
and divergent solutions to the Schr\"{o}dinger equation with the potential (%
\ref{eq:V_rational}) when $E=0$. It is worth noting that the RPM does not
distinguish between physical an unphysical results unless one manages to
obtain the wavefunction from its logarithmic derivative as we did it in this
example. In other cases, like the potential~(\ref{eq:V_MPT}), it may be
easier to resort to another approach to obtain the wavefunction from the
parameters given by the RPM.

In the appendix we solve the Schr\"{o}dinger equation for this potential and
derive the exact convergent and divergent solutions for $E=0$.

Although the RPM yields the exact critical parameters it is not
suitable for the calculation of the energies close to threshold.
The sequence of roots of $H_{D}^{d}(E,v_{0})=0$ converge rather
too slowly when $v_{0}$ is close (slightly greater than) a
critical parameter. When $E<0$ the function $f(x)$ is not an exact
rational function and the RPM yields approximately those
eigenvalues that are not too close to threshold. We calculated the
ground state for $v_{0}=3/2$ ($E_{1}=0$) and the first two bound
states for $v_{0}=4 $ ($E_{2}=0$). In the three cases we found
that the sequences $E^{[D,0]}$ converge from above and the
sequences $E^{[D,1]}$ from below. We assume that the former
provides upper bounds and the latter lower ones as in an earlier
treatment of the quartic anharmonic
oscillator\cite{FMT89a,FMT89b}. Thus, from sequences of roots with
$D\leq 80$ we conjecture that
\begin{eqnarray}
 -0.6985262171667534202327 &>&E_{0}>-0.6985262171667534202332,\,v_{0}=\frac{3}{2} ,  \nonumber \\
-2.4713450252412636948012742 &>&E_{0}>-2.4713450252412636948012743,\,v_{0}=4 ,  \nonumber \\
-0.42640598980647065078&>&E_{1}>-0.42640598980647065113,\,v_{0}=4.
\end{eqnarray}
We are not aware of any calculation of the eigenvalues and eigenfunctions
for this rational potential.

\section{Modified RPM for critical parameters}

\label{sec:MRPM}

According to the results of section~\ref{sec:wells} the appropriate boundary
condition at threshold is given by equation (\ref{eq:BC_E=0}). Therefore, it
seems reasonable to look for an ansatz with poles at the zeros of $\psi
^{\prime }(x)$. One suitable choice is the function
\begin{equation}
g(x)=\frac{1-s}{x}-\frac{\psi ^{\prime \prime }(x)}{\psi ^{\prime }(x)} .
\label{eq:g(x)}
\end{equation}
We thus have
\begin{equation}
\frac{1-s}{x}f(x)-f(x)g(x)+\frac{s}{x}g(x)=Q(E,x) ,  \label{eq:eq_for_g(x)}
\end{equation}
and
\begin{equation}
g(x)=x\sum_{j=0}^{\infty }g_{j}x^{2j} .  \label{eq:g(x)_series}
\end{equation}
The first coefficients are

\begin{eqnarray}
g_{0} &=&\frac{Q_{0}}{3}-\frac{2Q_{1}}{3Q_{0}} ,  \nonumber \\
g_{1} &=&\frac{Q_{0}^{2}}{45}-\frac{4Q_{2}}{5Q_{0}}+\frac{2Q_{1}^{2}}{%
9Q_{0}^{2}}+\frac{11Q_{1}}{45},
\end{eqnarray}
for $s=0$ and
\begin{eqnarray}
g_{0} &=&Q_{0} ,  \nonumber \\
g_{1} &=&Q_{1}+\frac{1}{3}Q_{0}^{2} ,  \nonumber \\
g2 &=&Q_{2}+\frac{8}{15}Q_{0}Q_{1}+\frac{2}{15}Q_{0}^{3},
\end{eqnarray}
for $s=1$. We apply the RPM exactly in the same way and construct the Hankel
determinants with the coefficients $g_{j}$: $H_{D}^{d}(E,v_{0})=\left|
g_{i+j+d-1}(E,v_{0})\right| _{i,j=1}^{D}=0$.

We obtain convergent sequences of roots of $H_{D}^{d}(E=0,v_{0})=0$ for all
the models discussed above. In particular, Table~\ref{tab:GW} shows the
first critical parameters for the Gaussian well estimated from the roots of
the Hankel determinants with $D\leq 80$, $d=0$ and $d=1$. For comparison we
add the results of Liverts and Barnea\cite{LB11} for the s-states of the
central-field model. The critical parameters for the central-field model
with angular momentum quantum number $l=0$ are exactly those for the odd
states of the one-dimensional case.

In closing this section we mention that we also tried the alternative ansatz
$\psi(x)/\psi^\prime(x)$ for odd eigenfunctions but in this case the rate of
convergence proved to be considerably lower.

\section{Central-field models}

\label{sec:central-field}

The results of section~\ref{sec:MRPM} suggest that the present approach may
also be suitable for the s-states of other central-field models. Although
the present paper is focused on one-dimensional parity-invariant models we
can outline a strategy for the treatment of central-field models. We write
the radial part of the dimensionless Schr\"{o}dinger equation as
\begin{eqnarray}
&&\psi ^{\prime \prime }(r)+\left[ Q(r)-\frac{l(l+1)}{r^{2}}\right] \psi
(r)=0,  \nonumber \\
&&Q(r)=2\left[ E-V(r)\right] ,\,\psi (0)=0,
\end{eqnarray}
and assume that
\begin{equation}
Q(r)=\sum_{j=-1}^{\infty }Q_{j}r^{j}.
\end{equation}
As in earlier papers we define\cite{FMT89a}
\begin{equation}
f(r)=\frac{l+1}{r}-\frac{\psi ^{\prime }(r)}{\psi (r)},
\end{equation}
and in order to apply the modified RPM to the calculation of critical
parameters we resort to the auxiliary function
\begin{equation}
g(r)=\frac{l}{r}-\frac{\psi ^{\prime \prime }(r)}{\psi ^{\prime }(r)}.
\end{equation}
They are related by
\begin{equation}
\frac{l+1}{r}g(r)-f(r)g(r)+\frac{l}{r}f(r)-Q(r)=0,
\end{equation}
and can be expanded in a Taylor series about the origin as
\begin{eqnarray}
f(r) &=&\sum_{j=0}^{\infty }f_{j}r^{j},  \nonumber \\
g(r) &=&\sum_{j=0}^{\infty }g_{j}r^{j}.
\end{eqnarray}
As an illustrative example we choose the Yukawa potential
\begin{equation}
V(r)=-\frac{v_{0}e^{-r}}{r},  \label{eq:V_Yuk}
\end{equation}
and show the results in Table~\ref{tab:Yuk_s} for the first s-states
estimated from roots of the $g$-Hankel determinants with $D\leq 80$, $d=0$
and $d=1$. Present results agree with those of Liverts and Barnea\cite{LB11}
and Singh and Varshni\cite{SV84} up to the last digit reported by them.

The asymptotic behaviour of the solutions to the central-field models when $%
E=0$ is given by
\begin{equation}
\psi (r)\sim Ar^{-l}+Br^{l+1}.
\end{equation}
Therefore, we expect that the original RPM yields reasonable critical
parameters for $l>0$. In other words, the roots of the $f$-Hankel
determinants are expected to approach the roots of $B(v_{0,n,l})=0$ as the
determinant dimension increases. Tables \ref{tab:Yuk_l} and \ref{tab:Gaus_l}
clearly show  that the rate of convergence of the RPM increases with $l$ and
decreases with $n$.

Since the accuracy of the RPM increases with $l$ we can test the WKB large-$%
l $ asymptotics $\beta _{n,l}\sim el(l+1)$ derived by Liverts and Barnea\cite
{LB11} for both the Yukawa and Gaussian potentials (note that $\beta
_{n,l}=2v_{0,n,l}$). We can also compare these results with the variational
estimates
\begin{equation}
v_{0,1,l}^{Y}=\frac{2^{2l}\left( l+1\right) ^{2l+3}}{\left( 2l+1\right)
^{2l+1}},
\end{equation}
and
\begin{equation}
v_{0,1,l}^{G}=\frac{\left( 2l+3\right) ^{\frac{2l+5}{2}}}{8\left(
2l+1\right) ^{\frac{2l+1}{2}}},
\end{equation}
derived by means of the trial functions $\varphi (r)=Nr^{l+1}e^{-ar}$ and $%
\varphi (r)=Nr^{l+1}e^{-ar^{2}}$ for the Yukawa and Gaussian potentials,
respectively\cite{FG13}.

Tables \ref{tab:Yuk_large_l} and \ref{tab:Gauss_large_l} show the RPM, WKB
and variational results, as well as the logarithmic errors of the two latter
ones. We appreciate that the variational estimates are somewhat more
accurate but the WKB expression shows the striking fact that the large-$l$
asymptotic behaviour for the critical parameters for both potentials is
exactly the same. Although the two variational results are different for
small and moderate $l$ they agree with the WKB ones for sufficiently large $l
$:
\begin{equation}
\lim\limits_{l\rightarrow \infty }\frac{v_{0,1,l}^{Y}}{l(l+1)}%
=\lim\limits_{l\rightarrow \infty
}\frac{v_{0,1,l}^{G}}{l(l+1)}=\frac{e}{2}.
\end{equation}
The RPM results in tables \ref{tab:Yuk_large_l} and \ref{tab:Gauss_large_l}
are accurate to the last digit and were obtained by means of Hankel
determinants of dimension as small as $D=10$.

The rate of convergence of the  modified RPM based on $g$-Hankel
determinants also increases with $l$ but we do not deem necessary to show
those results.

Although present results are more accurate than those of Liverts and Barnea%
\cite{LB11} and Singh and Varshni\cite{SV84} one should not conclude that
the RPM is superior to the approaches developed by those authors. Those
other methods are more general because present local approximation is based
on the Taylor expansion of the solution about some chosen point which limits
the class of potentials that can be treated successfully. However, the RPM
is a straightforward simple approach that applies to a wide variety of
problems. In many cases it yields quite accurate results and may be suitable
for testing other approaches and even for setting benchmark data.

\section{Perturbation theory about the threshold}

\label{sec:PT}

We can expand the exact energy (\ref{eq:En_MPT}) for the modified
P\"{o}schl-Teller potential in a Taylor series about $v_{0,n}$ and obtain
the perturbation series about the threshold
\begin{eqnarray}
E_{n} &=&-\frac{2\xi ^{2}}{\left( 2n+1\right) ^{2}}+\frac{8\xi ^{3}}{\left(
2n+1\right) ^{4}}-\frac{40\xi ^{4}}{\left( 2n+1\right) ^{6}}+\frac{224\xi
^{5}}{\left( 2n+1\right) ^{8}}+O(\xi ^{6}) ,  \nonumber \\
\xi &=&v_{0}-v_{0,n} ,  \label{eq:E_MPT_PT}
\end{eqnarray}
that converges for all $|\xi |<(2n+1)^{2}/8$. Note that the perturbation
correction of first order is zero for all states and that we obtain a
negative energy for both $v_{0}>v_{0,n}$ and $v_{0}<v_{0,n}$ if $\xi $ is
sufficiently small, in spite of the fact that the $n$-th state moves into
the continuum in the latter case. We can carry out a similar calculation for
models that are not exactly solvable by means of the RPM. In what follows we
illustrate the strategy by means of the apparently most favourable case of
the rational potential (\ref{eq:V_rational}) for which the RPM yields the
exact solution at threshold.

The roots of a Hankel determinant $H_{D}^{d}(E,v_{0})=0$ give us
approximations to either $E(v_{0})$ or $v_{0}(E)$. If we substitute
\begin{equation}
E=E^{(1)}\xi +E^{(2)}\xi ^{2}+\ldots +E^{(k)}\xi ^{k} ,  \label{eq:E_PT}
\end{equation}
and $v_{0}=v_{0,n}+\xi $ into the Hankel determinant, then we can obtain the
coefficients $E^{(j)}$, $j=1,2,\ldots k$ of the perturbation series, the
accuracy increasing with $D$. Based on the Hellmann-Feynman theorem\cite{F39}
(see also\cite{F04} for a discussion about degenerate states)
\begin{equation}
\frac{dE}{dv_{0}}=-\left\langle \frac{1}{\left( 1+x^{2}\right) ^{2}}%
\right\rangle ,  \label{eq:H-F_theorem}
\end{equation}
we expect that
\begin{equation}
E^{(1)}=\lim_{\xi \rightarrow 0^{+}}\frac{dE}{d\xi }\leq 0 ,
\label{eq:E(1)_def}
\end{equation}
for a physically acceptable solution. Since the solutions are not square
integrable when $\xi =0$ then the expectation value in equation (\ref
{eq:H-F_theorem}) is meaningless at threshold; however the limit (\ref
{eq:E(1)_def}) may hopefully be finite. In fact, $E^{(1)}=0$ for the
P\"{o}schl-Teller potential.

It follows from the discussion in the subsection~\ref{subsec:RatPot} that $%
E_{1}=0$ when $v_{0}=3/2$. However, if we substitute $v_{0}=3/2+\xi $ and
the series (\ref{eq:E_PT}) into the Hankel determinants for $s=1$ we obtain
the unphysical result
\begin{equation}
E=\frac{1}{8}\xi -\frac{7}{64}\xi ^{2}+\frac{29}{768}\xi ^{3}-\frac{1847}{%
184320}\xi ^{4}+\frac{275357}{77414400}\xi ^{5}+O(\xi ^{6}) .
\label{eq:E_PT_n=1_s=1}
\end{equation}
According to this expansion the energy increases as $v_{0}$ increases beyond
$v_{0,1}=3/2$ in contradiction with (\ref{eq:H-F_theorem}) and (\ref
{eq:E(1)_def}). This result reflects the fact mentioned above that the RPM
does not yield the energy $E_{1}$ for $v_{0}$ close to threshold.

If we repeat the calculation for the even states we obtain a perturbation
expansion with the expected slope at threshold:

\begin{equation}
E=-\frac{1}{8}\xi +\frac{17}{192}\xi ^{2}-\frac{23}{11520}\xi ^{3}-\frac{%
271933}{19353600}\xi ^{4}+\frac{29363423}{8128512000}\xi ^{5}+O(\xi ^{6}) .
\label{eq:E_PT_n=1_s=0}
\end{equation}
At first sight, this result is surprising because no new even state should
appear when $3/2<v_{0}<4$ (the ground state remains bound for all $v_{0}>0$%
). The explanation is that the RPM yields a spurious even-state energy
associated to the divergent solution $\psi _{1,0}$. For example, when $%
v_{0}=1.51$ the RPM with $D=10$ and $d=0$ yields the actual ground-state
energy $E_{0}\approx -0.70483$ and the spurious root $W\approx
-0.00124114797000675832$ . The considerably greater accuracy of the latter
is due to the fact that the RPM yields the exact result when $v_{0}=3/2$.
The question remains why the RPM does not yield the energy $E_{1}$ of the
odd state in view of the fact that the calculation of the critical parameter
is also exact in this case. The perturbation expansion (\ref{eq:E_PT_n=1_s=1}%
) with the wrong slope at threshold also describes a spurious root. For
example, when $v_{0}=1.49$ the roots of the Hankel determinants with $D=10$
and $d=0$ yields $W\approx -0.0012609753609799139195$ ($s=1$) and $%
E_{0}\approx -0.692231$ ($s=0$). It is clear that the RPM favours the
unphysical solutions; in fact, it is also interesting that the spurious
roots of the Hankel determinants follow the unphysical expansions (\ref
{eq:E_PT_n=1_s=0}) and (\ref{eq:E_PT_n=1_s=1}) for both $\xi <0$ and $\xi >0$
and in either case the rate of convergence of the corresponding sequences is
remarkably large. For example, the expansion~(\ref{eq:E_PT_n=1_s=0})
predicts a positive root for a negative value of $\xi $ and the RPM already
yields it quite accurately. When $v_{0}=1.49$ we obtain $%
W=0.0012588560223263235359$ in agreement with that perturbation series.

The second excited state vanishes when $v_{0}=4$. The RPM perturbation
expansions obtained by substitution of $v_{0}=4+\xi $ are also unphysical.
For example when $s=0$ we obtain a series with the wrong slope at threshold:

\begin{equation}
E=\frac{1}{32}\xi -\frac{23}{2304}\xi ^{2}-\frac{919}{331776}\xi ^{3}+\frac{%
100843}{59719680}\xi ^{4}-\frac{418250431}{1203948748800}\xi ^{5}+O(\xi
^{6}) .  \label{eq:E_PT_n=2_s=0}
\end{equation}
This results is not surprising if we take into account that the RPM fails to
give us the second excited state when $v_{0}$ is slightly larger than $4$.
When $s=1$ we obtain the exact power series for another unphysical root of
the Hankel determinants
\begin{equation}
E=-\frac{1}{32}\xi +\frac{7}{768}\xi ^{2}+\frac{1921}{552960}\xi ^{3}-\frac{%
1186027}{696729600}\xi ^{4}+\frac{2551967839}{14046068736000}\xi ^{5}+O(\xi
^{6}) .  \label{eq:E_PT_n=2_s=1}
\end{equation}
From the roots of the Hankel determinants for $v_{0}=4.01$ we obtain $%
E_{1}\approx -0.429395$ (the actual energy for the first-excited state) and
the spurious eigenvalue $W\approx -0.00031158508464057747545$ which is
associated to the divergent function $\psi _{2,1}$ when $\xi \rightarrow 0$ (%
$E\rightarrow 0^{-}$).

For some unknown reason the RPM yields the unphysical roots associated to
the divergent states more accurately than the physical ones stemming from
the convergent states. However, the approach is still a useful tool for
obtaining the eigenvalues and critical parameters of one-dimensional wells
as already shown above. The modified RPM discussed in section~\ref{sec:MRPM}
also yields the same spurious roots; therefore, we may conclude that such an
unexpected behaviour is inherent in the Hankel determinants constructed from
either the coefficients $f_{j}$ or $g_{j}$.

In an attempt to understand the baffling results discussed above we analyzed
the zeroes of the denominators of the Pad\'e approximants for energies in
the neighborhoods of the actual eigenvalues and the spurious ones. However,
we could not derive any reasonable rule from such study.

Although the RPM yields the whole set of critical values with either the
even or odd functions for the modified P\"{o}schl-Teller potential, the
Hankel determinants do not exhibit spurious roots in this case.

\section{Conclusions}

\label{sec:conclusions}

Despite of being a local approximation the RPM may be a useful tool for the
calculation of critical parameters of one-dimensional and central-field
quantum wells. In some cases, like the modified P\"{o}schl-Teller and
rational potentials, the original version of the approach yields accurate
results. In other cases, like the Gaussian potential, it is necessary to
resort to a modified algorithm that applies to one-dimensional models as
well as to the s-states of central-field ones. The accuracy of the estimated
critical parameters is satisfactory for the Gaussian and Yukawa potentials.
The original RPM proves to be suitable for states with $l>0$ and its
accuracy increases with this quantum number. For this reason the RPM appears
to be a powerful tool for the calculation of critical parameters for
extremely large values of the angular-momentum quantum number.

In the case of the modified P\"{o}schl-Teller and rational potentials the
RPM yields the whole set of critical parameters for both the even and odd
solutions to the Schr\"{o}dinger equation. Half of them are associated to
the divergent solutions. The occurrence of unphysical results is due to the
fact that the RPM does not take into account the asymptotic behaviour of the
eigenfunctions explicitly. For some unknown reason the RPM seems to favour
the divergent solutions in the case of the rational potential. This
undesirable behaviour is not a serious limitation because there is no doubt
about which roots are spurious. We have not been able to give a sound answer
to this anomalous behavior from the roots of the denominator of the Pad\'e
approximants.

The P\"{o}schl-Teller and rational potentials are different in the sense
that the RPM yields accurate energies close to the threshold in the former
case but not in the latter one. Therefore, the three one-dimensional
potentials discussed in section~\ref{sec:examples} reveal three different
behaviour patterns in the application of the RPM to simple one-dimensional
parity-invariant quantum wells.

The RPM yields the exact solutions for the rational potential (\ref
{eq:V_rational}) when $E=0$. This particularly fortunate situation enables
one to try perturbation theory about the threshold. Surprisingly, the RPM
yields exact perturbation series only for the unphysical case of divergent
functions. The reason may be that the physically meaningful solutions do not
exhibit power-series expansions about threshold. However, we know that such
expansions exist in the case of the P\"{o}schl-Teller potential as shown in
Eq.~(\ref{eq:E_MPT_PT}). In this case the Hankel determinants do not exhibit
spurious roots.

We have also carried out calculations for potentials of the form $%
V(x)=-v_{0}/(1+x^{2})^{m}$, where $m=5/2,3,4$. In these cases the RPM fails
to provide the critical parameters and the modified RPM exhibits convergent
roots. The even and odd critical parameters appear separately as in the case
of the Gaussian potential discussed in subsection~\ref{subsec:Gaussian}.

\section{Appendix}

If we change the independent and dependent variables in the Schr\"{o}dinger
equation with the rational potential (\ref{eq:V_rational}) according to
\begin{eqnarray}
x &=&iz ,  \nonumber \\
\varphi (z) &=&\psi (iz)=\sqrt{1-z^{2}}w(z),
\end{eqnarray}
then we obtain
\begin{equation}
\frac{d}{dz}\left( 1-z^{2}\right) \frac{d}{dz}w+\left[ -2E\left(
1-z^{2}\right) -\frac{2v_{0}+1}{1-z^{2}}\right] w=0 .  \label{eq:Schro_w(z)}
\end{equation}
This equation is a particular case of the spheroidal differential equation
\begin{equation}
\frac{d}{dz}\left( 1-z^{2}\right) \frac{d}{dz}w+\left[ \lambda +\gamma
^{2}\left( 1-z^{2}\right) -\frac{\mu ^{2}}{1-z^{2}}\right] w=0,
\end{equation}
with $\lambda =0$, $\gamma ^{2}=-2E$ and $\mu ^{2}=2v_{0}+1$. We can also
relate equation (\ref{eq:Schro_w(z)}) with the associated Legendre equation
\begin{equation}
\frac{d}{dz}\left( 1-z^{2}\right) \frac{d}{dz}w+\left[ \nu (\nu +1)-\frac{%
\mu ^{2}}{1-z^{2}}\right] w=0,
\end{equation}
when $E=0$ and $\nu =0$.

However, for the present discussion we prefer to proceed in a different way.
The rational potential (\ref{eq:V_rational}) exhibits singularities at $%
x=\pm i$. If we substitute
\begin{equation}
\psi (x)=\left( 1+x^{2}\right) ^{\alpha }u(x),
\end{equation}
into the Schr\"{o}dinger equation we obtain an eigenvalue equation for the
new dependent variable $u(x)$:
\begin{eqnarray}
\left( 1+x^{2}\right) u^{\prime \prime }+4\alpha xu^{\prime }-\frac{2\left(
2\alpha ^{2}-2\alpha -v_{0}\right) }{1+x^{2}}u &&  \nonumber \\
+\left[ 2Ex^{2}+2\left( 2\alpha ^{2}-\alpha +E\right) \right] u &=&0.
\end{eqnarray}
We remove the third term by choosing $\alpha $ to be any one of the roots of
\begin{equation}
2\alpha ^{2}-2\alpha -v_{0}=0.
\end{equation}

Then, we expand $u$ in a Taylor series about the origin
\begin{equation}
u(x)=\sum_{j=0}^{\infty }c_{j}x^{2j+s},
\end{equation}
and derive a three-term recurrence relation for the coefficients
\begin{eqnarray}
(2j+s+1)(2j+s+2)c_{j+1}&+&\left[ (2j+s+2\alpha )(2j+s+2\alpha
-1)+2E\right]c_{j}  \nonumber \\
&+&2Ec_{j-1}=0.
\end{eqnarray}
The radius of convergence of this series is unity $\left( \lim_{j\rightarrow
\infty }\left| \frac{c_{j+1}}{c_{j}}\right| =1\right) $ because of the
singularities at $x=\pm i$. When $E=0$ the recurrence relation becomes a
two-term one and we can obtain polynomial solutions for particular values of
$\alpha $. Note that if
\begin{equation}
\alpha =\left\{
\begin{array}{l}
\alpha _{1}(k,s)=-k-\frac{s}{2} \\
\alpha _{2}(k,s)=-k-\frac{s}{2}+1
\end{array}
\right. ,
\end{equation}
then $c_{j}=0$ for all $j>k$.We thus have two sets of critical potential
parameters:
\begin{equation}
v_{0}=\frac{(2\alpha -1)^{2}-1}{2}=\left\{
\begin{array}{l}
v_{0}^{(1)}(k,s)=\frac{(2k+s-1)^{2}-1}{2} \\
v_{0}^{(2)}(k,s)=\frac{(2k+s)^{2}-1}{2}
\end{array}
\right. ,
\end{equation}
where $v_{0}^{(1,2)}=v_{0}(\alpha _{1,2})$. It is interesting to note that
the critical parameters exhibit a kind of degeneracy:
\begin{eqnarray}
v_{0}^{(1)}(k,1) &=&v_{0}^{(2)}(k,0)=\frac{4k^{2}-1}{2},\,k=1,2,\ldots ,
\nonumber \\
v_{0}^{(1)}(k+1,0) &=&v_{0}^{(2)}(k,1)=\frac{(2k+1)^{2}-1}{2},\,k=1,2,\ldots,
\end{eqnarray}
similar to the one found by Joseph\cite{J67b,J67a} for the central-field
version of this model. In the present case one of the degenerate solutions
is convergent ($\sim 1$) and the other one is divergent ($\sim x$). The
connection between both models becomes apparent if we take into account that
the states of the central-field model with angular-momentum quantum numbers $%
l=-1$ and $l=0$ become the even and odd states of the one-dimensional one.

In order to understand the results derived in the subsection~\ref
{subsec:RatPot} by means of the RPM simply note that
\begin{equation}
f(x)=\frac{s}{x}-\frac{\psi ^{\prime }(x)}{\psi (x)}=\frac{s}{x}-\frac{%
2\alpha x}{1+x^{2}}-\frac{u^{\prime }(x)}{u(x)},
\end{equation}
is a rational function of $x$.

\begin{table}[]
\caption{Critical parameters for the Gaussian Well}
\label{tab:GW}
\begin{center}
\par
{\tiny
\begin{tabular}{|D{.}{.}{2}|D{.}{.}{20}|D{.}{.}{7}|}
\hline \multicolumn{1}{|c}{n} & \multicolumn{1}{|c|}{$v_{0,n}$
(present)} & \multicolumn{1}{c|}{$\beta/2$ \cite{LB11}}
\\ \hline
1 & 1.34200232546204576914 & 1.3420023 \\
2 & 4.32454875170105636793 &\\
3 & 8.89784977356695359410 & 8.89785 \\
4 & 15.05314025436583553157 &\\
5 & 22.78673996005213242180 & 22.78674 \\
6 & 32.09666656038554309293 &\\
7 & 42.98170019005867752947 & 42.9817 \\
8 & 55.44102390556364979485 &\\
9 & 69.47405735384177416019 & 69.47406 \\
10 & 85.08036985819273906133 &\\
11 & 102.2596308675957148370 & 102.25963 \\
12 & 121.0115797852355989558 &\\
13 & 141.3360066230547124163 & 141.33601 \\
14 & 163.2327390694287387382 &\\
15 & 186.7016335410971417677 & 186.70163 \\
16 & 211.7425688100812690936 &\\
17 & 238.3554413511863843581 & 238.35544 \\
18 & 266.5401618724453436776 &\\
19 & 296.2966526792028587502 & 296.29665 \\
20 & 327.6248456385162999270 &\\
21 & 360.5246805841777739108 & 360.52468 \\
\hline
\end{tabular}
}
\end{center}
\end{table}

\begin{table}[]
\caption{Critical parameters for the Yukawa potential: s states}
\label{tab:Yuk_s}
\begin{center}
\begin{tabular}{|D{.}{.}{2}|D{.}{.}{26}|D{.}{.}{8}|D{.}{.}{15}|}
\hline \multicolumn{1}{|c}{State} & \multicolumn{1}{|c|}{$v_{0,n}$
(present)} & \multicolumn{1}{|c|}{$\beta/2$ \cite{LB11}} &
\multicolumn{1}{c|}{$1/\delta$ \cite{SV84}} \\ \hline
1s & 0.83990388669822801527775556 & 0.83990390 & 0.839903886698226 \\
2s & 3.2236301610682666483973 & 3.2236302 & 3.22363017 \\
3s & 7.17101392084392858317 & 7.1710140 & 7.17101392 \\
4s & 12.68582992202390726756 & 12.685830 & 12.685830 \\
5s & 19.76942118485633321537 & 19.769421 & 19.769421 \\
6s & 28.42243219866087345719 & 28.42243 & 28.422432 \\
7s & 38.64522743052775121132 & 38.645227 & 38.645227 \\\hline
\end{tabular}
\end{center}
\end{table}

\begin{table}[]
\caption{Critical parameters for the Yukawa potential: states with $l>0$ }
\label{tab:Yuk_l}
\begin{center}
\begin{tabular}{|D{.}{.}{1}|D{.}{.}{1}|D{.}{.}{21}|D{.}{.}{7}|D{.}{.}{9}|}
\hline \multicolumn{1}{|c}{$l$} & \multicolumn{1}{|c|}{$n$} &
\multicolumn{1}{|c|}{RPM} & \multicolumn{1}{c|}{$\beta_{n,l} / 2$
\cite{LB11}} & \multicolumn{1}{c|}{$1/\delta$ \cite{SV84} }
\\ \hline
1 & 1 & 4.540979480 & 4.5409795              &  4.540979547   \\
1 & 2 & 8.872287943 & 8.872288               &  8.87228793  \\
1 & 3 & 14.7307131 & 14.730713               &  14.730713  \\
1 & 4 & 22.1306205 & 22.130627               &  22.130627  \\
2 & 1 & 10.947492231128 & 10.947492          &  10.947492  \\
2 & 2 & 17.21020724698 & 17.210207           &  17.210207 \\
2 & 3 & 24.98478805031 & 24.984788           &  24.984788 \\
2 & 4 & 34.285733608 & 34.2857335            &  34.285734  \\
3 & 1 & 20.06777597598021672 & 20.067776     &  20.067776  \\
3 & 2 & 28.257056865147125 & 28.257057       &  28.257057 \\
3 & 3 & 37.949696830060 & 37.949697          &  37.949696 \\
3 & 4 & 49.1589622686 & 49.1589625           &  49.158964\\
4 & 1 & 31.904488236447390251 & 31.904488    &  31.904488 \\
4 & 2 & 42.01838864622171175 & 42.0183885    &  42.018390 \\
4 & 3 & 53.6301861108720125 & 53.630185      &  53.630187 \\
4 & 4 & 66.7518302698487 & 66.75183          &  66.751829 \\
5 & 1 & 46.458582142052657720625 & 46.458582 &  46.458583 \\
5 & 2 & 58.4961723904053473472 & 58.49617    &  58.496171 \\
5 & 3 & 72.02784452443215966 & 72.027845     &  72.027848\\
5 & 4 & 87.0643772674642755 & 87.064375      &  \\

\hline
\end{tabular}
\end{center}
\end{table}

\begin{table}[]
\caption{Critical parameters for the gaussian well: states with $l>0$}
\label{tab:Gaus_l}
\begin{center}
\begin{tabular}{|D{.}{.}{1}|D{.}{.}{1}|D{.}{.}{12}|D{.}{.}{8}|}
\hline \multicolumn{1}{|c}{$l$} & \multicolumn{1}{|c|}{$n$} &
\multicolumn{1}{c|}{RPM}&\multicolumn{1}{c|}{$\beta_{n,l}/2$
\cite{LB11}} \\ \hline
1 & 1 & 6.049654263     & 6.0496545   \\
1 & 2 & 17.544888       & 17.544888   \\
1 & 3 & 35.241431       & 35.241428  \\
1 & 4 & 59.17581        & 59.175825  \\
2 & 1 & 13.4505387996   & 13.450539  \\
2 & 2 & 28.83788607     & 28.837886  \\
2 & 3 & 50.35752508     & 50.357525  \\
2 & 4 & 78.063746       & 78.063745 \\
3 & 1 & 23.553930851605 & 23.553931   \\
3 & 2 & 42.81369669354  & 42.813696   \\
3 & 3 & 68.162501708    & 68.162500  \\
3 & 4 & 99.65923348     & 99.659230 \\
4 & 1 & 36.366501836074 & 36.366502   \\
4 & 2 & 59.48855740034  & 59.488555  \\
4 & 3 & 88.6698082860   & 88.669810   \\
4 & 4 & 123.9690574563  & 123.96905   \\\hline
\end{tabular}
\end{center}
\end{table}

\begin{table}[]
\caption{Critical parameters for the Yukawa potential with $n=1$ and large $%
l $ calculated by means of the RPM, WKB and variational approaches. LE is
the logarithmic error: $\log|\left(\mathrm{{exact}-{approximate}}\right)/%
\mathrm{{exact}|}$}
\label{tab:Yuk_large_l}
\begin{center}
\begin{tabular}{|D{.}{.}{1}|D{.}{.}{16}|D{.}{.}{2}|D{.}{.}{3}|D{.}{.}{3}|
D{.}{.}{3}|}
\hline \multicolumn{1}{|c}{$l$} & \multicolumn{1}{|c|}{RPM} &
\multicolumn{1}{|c|}{WKB\cite{LB11}} & \multicolumn{1}{c|}{LE} & \multicolumn{1}{c|}{Variational }
& \multicolumn{1}{c|}{LE }\\ \hline

50  &   3514.7478136194717430  &   3466    &    -1.9  &     3518    &     -3.1 \\
100 &   13824.314996806666238  &   13727   &    -2.2  &     13830   &     -3.4 \\
150 &   30929.586489910790437  &   30785   &    -2.3  &     30938   &     -3.5 \\
200 &   54830.562488458876745  &   54637   &    -2.5  &     54842   &     -3.7 \\
250 &   85527.243031809696017  &   85286   &    -2.5  &     85542   &     -3.8 \\
300 &   123019.62813311806352  &   122730  &    -2.6  &     123037  &     -3.8 \\
350 &   167307.71779803038796  &   166970  &    -2.7  &     167328  &     -3.9 \\
400 &   218391.51202937270652  &   218006  &    -2.8  &     218415  &     -4.0 \\
450 &   276271.01082871615922  &   275838  &    -2.8  &     276297  &     -4.0 \\
500 &   340946.21419700393386  &   340465  &    -2.9  &     340975  &     -4.1 \\

\hline
\end{tabular}
\end{center}
\end{table}

\begin{table}[]
\caption{Critical parameters for the Gaussian potential with $n=1$ and large
$l$ calculated by means of the RPM, WKB and variational approaches. LE is
the logarithmic error: $\log|\left(\mathrm{{exact}-{approximate}}\right)/%
\mathrm{{exact}|}$ }
\label{tab:Gauss_large_l}
\begin{center}
\begin{tabular}{|D{.}{.}{1}|D{.}{.}{16}|D{.}{.}{2}|D{.}{.}{3}|D{.}{.}{3}|
D{.}{.}{3}|} \hline \multicolumn{1}{|c}{$l$} &
\multicolumn{1}{|c|}{RPM} & \multicolumn{1}{|c|}{WKB\cite{LB11}} &
\multicolumn{1}{c|}{LE} & \multicolumn{1}{c|}{Variational } &
\multicolumn{1}{c|}{LE }\\ \hline

50  &  3563.4739040116520856    & 3466   & -1.6   &  3570    &  -2.8  \\
100 &   13921.096733881441676   &  13727 &  -1.9  &   13933  &   -3.1  \\
150 &   31074.422038210807820   &  30785 &  -2.0  &   31092  &   -3.2  \\
200 &   55023.451386983337338   &  54637 &  -2.2  &   55047  &   -3.4  \\
250 &   85768.185095826709503   &  85286 &  -2.3  &   85798  &   -3.5  \\
300 &   123308.62327017069510   &  122730&  -2.3  &   123344 &   -3.5  \\
350 &   167644.76595525290586   &  166970&  -2.4  &   167686 &   -3.6  \\
400 &   218776.61317370956855   &  218006&  -2.5  &   218824 &   -3.7  \\
450 &   276704.16493812321886   &  275838&  -2.5  &   276757 &   -3.7  \\
500 &   341427.42125604645356   &  340465&  -2.5  &   341486 &   -3.8  \\

\hline
\end{tabular}
\end{center}
\end{table}

\end{document}